# Transport critical current density in Fe-sheathed nano-SiC doped MgB$_2$ wires

Shi X. Dou, Joseph Horvat, Saeid Soltanian, Xiao L. Wang, Meng J. Qin, Shi H. Zhou, Hua K. Liu and Paul G Munroe

*Abstract*—The nano-SiC doped MgB$_2$/Fe wires were fabricated using a powder-in-tube method and an in-situ reaction process. The depression of T$_c$ with increasing SiC doping level remained rather small due to the counterbalanced effect of Si and C co-doping. The high level SiC co-doping allowed creation of the intra-grain defects and nano-inclusions, which act as effective pinning centers, resulting in a substantial enhancement in the J$_c$(H) performance. The transport J$_c$ for all the wires is comparable to the magnetic J$_c$ at higher fields despite the low density of the samples and percolative nature of current. The transport I$_c$ for the 10wt% SiC doped MgB$_2$/Fe reached 660A at 5K and 4.5T (J$_c$ = 133,000A/cm$^2$) and 540A at 20K and 2T (J$_c$ = 108,000A/cm$^2$). The transport J$_c$ for the 10wt% SiC doped MgB$_2$ wire is more than an order of magnitude higher than for the state-the-art Fe-sheathed MgB$_2$ wire reported to date at 5K and 10T and 20K and 5T respectively. There is a plenty of room for further improvement in J$_c$ as the density of the current samples is only 50%.

*Index Terms*—Critical current, nano-particle doping, magnesium diboride, silicon carbide.

## I. INTRODUCTION

It has been established that Fe sheath is suitable for fabrication of MgB$_2$ wires using a powder-in-tube method [1], [2]. Extensive research efforts have been made to improve the J$_c$ of Fe-sheathed MgB$_2$ wires [1]-[5]. However, the J$_c$ performance in high fields and temperatures above 20K remains unsatisfactory for many appliations due to the poor pinning ability of this material. Attempts to enhance flux pinning have resulted in an encouraging improvement in irreversibility fields (H$_{irr}$) and J$_c$(H) by oxygen alloying in MgB$_2$ thin films [6] and by proton irradiation of MgB$_2$ powder [7]. However, these techniques are not readily available for introducing effective pinning centers into MgB$_2$ wires. Chemical doping has been commonly used to introduce flux pinning centers into a superconductor for enhancing critical current density. Unfortunately, chemical doping of MgB$_2$ reported so far is limited to addition, rather than substitution of the elements. The additives alone appear to be ineffective for improving pinning at high temperatures [8-10].

Recently, we found that chemical doping with nano-particle SiC into MgB$_2$ can significantly enhance J$_c$ in high fields with only slight reduction in T$_c$ up to the doping level of 40% of B [11]. This finding clearly demonstrated that co-substitution of SiC for B in MgB$_2$ induced intra-grain defects and high density of nano-inclusions as effective pinning centers, responsible for the improved performance of J$_c$(H) in wide range of temperatures [12]. However, all the results reported thus far have been limited to magnetic measurements. As the materials are far from optimum and the sample density was only 50% of theoretical value the current in such a porous material is highly percolative. The major concern is whether the material can carry large transport J$_c$. In this work, we focus our study on the transport current and its field dependence for the nanometer-size SiC doped MgB$_2$ wires. Our results reveal that the nanometer size SiC doped MgB$_2$/Fe wires can carry higher transport I$_c$ and J$_c$ in the magnetic fields ever reported for any form of MgB$_2$. SiC doped MgB$_2$ is very promising for many applications, as this chemical doping is a readily achievable and economically viable process to introduce effective flux pinning.

## II. EXPERIMENTAL DETAILS

A standard powder-in-tube method was used for the Fe clad MgB$_2$ tape [2]. Powders of magnesium (99%) and amorphous boron (99%) were well mixed with 0 and 10 wt% of SiC nano-particle powder (size of 10nm to 20nm) and thoroughly ground. The pure Fe tube had an outside diameter (OD) of 10 mm, a wall thickness of 1 mm, and was 10 cm long with one end of the tube sealed. The mixed powder was filled in to the tube and the remaining end was crimped by hand. The composite was drawn to a 1.4mm diameter wire 2 meters long. Several short samples 2 cm in length were cut from the wire. These pieces were then sintered in a tube furnace over a temperature range from 800-850$^o$C for 10min to 30min. This was followed by furnace cooling to room temperature. A high purity argon gas flow was maintained throughout the sintering process.

This work was supported in part by the Australia Research Council, Hyper Tech Research Inc., U.S. and Alphatech International Ltd

S.X. Dou is with the Institute for Superconducting and Electronic Materials, University of Wollongong, Northfields Ave. Wollongong, NSW 2522 Australia (the corresponding author phone: 61-2-4221-4558; fax: 61-2-4221-5731; e-mail: shi_dou@uow.edu.au).
J. Horvat is with University of Wollongong (e-mail: jhorvat@uow.edu.au).
S. Soltanian is with University of Wollongong (e-mail: ss27@uow.edu.au).
X.L.Wang is with University of Wollongong(e-mail:xiaolin@uow.edu.au).
M.J. Qin is with University of Wollongong (e-mail:qin@uow.edu.au)
S.H. Zhou is with University of Wollongong (e-mail: sz03@uow.edu.au).
H.K. Liu is with University of Wollongong (e-mail: hua_liu@uow.edu.au).
P.G. Munroe is with the Electron Microscope Unit, University of New South Wales, NSW 2001 Australia (e-mail: p.munroe@unsw.edu.au)

Transport current was measured using pulse DC method. A pulse of the current was obtained by discharging a capacitor through the sample, coil of thick copper wire and non-inductive resistor connected in series. The current was measured via the voltage drop on the non-inductive resistor of 0.01 Ohm. The current reached its maximum value (700A) within 1ms. The voltage developed on the sample was measured simultaneously with the current, using a 2-channel digital oscilloscope. Because both channels of the oscilloscope had the same ground, the signal from the voltage taps was first fed to a transformer preamplifier (SR554). This decoupled the voltage taps from the resistor used for measuring the current, thereby avoiding creation of the ground loops and parasitic voltages in the system, as well as of an additional current path in parallel to the sample. The transformer amplified the voltage 100 times, improving the sensitivity of the experiment. Magnetic field was produced by a 12T superconducting magnet. Sample mounting allowed for orienting the field either perpendicular to the wire, or parallel to it. In the later case, the field was also parallel to the current passing through the sample. The sample was placed into a continuous flow helium cryostat, allowing the control of temperature better than 0.1K.

The magnetization of samples was measured over a temperature range of 5 to 30 K using a Physical Property Measurement System (PPMS, Quantum Design) with a sweep rate of magnetic field of 50 Oe/s and amplitude up to 8.5T. Samples are in the form of bars cut from the pellets which were processed under the same conditions as the wires. All the samples had the same size of 0.56x2.17x3.73 mm$^3$. A magnetic $J_c$ was derived from the height of the magnetization loop using Bean's model.

## III. Results and Discussion

Fig. 1 shows the transition temperature ($T_c$) for the doped and undoped samples determined by ac susceptibility measurements. The $T_c$ obtained as the onset of magnetic screening for the undoped sample was 37.6K. For the 10wt% SiC doped sample, the $T_c$ decreased only for 0.7K. In contrast, the $T_c$ was depressed by almost 7K for 10% C substitution for B in MgB$_2$ [13]. This suggests that the higher tolerance of $T_c$ of MgB$_2$ to SiC doping is attributable to the counterbalance effect of the co-doping by C and Si. This is because the average size of C (0.077nm) and Si (0.11nm) atoms is close to that of B (0.097nm). As reported previously, the 10wt% SiC doped sample consists of major phase with MgB$_2$ structure and minority phases: Mg$_2$Si and MgO which occupied about 10% to 15% volume fraction.

Fig. 2 shows a typical V-I characteristic for the MgB$_2$/Fe wire. It is noted that the self-field of the current pulse induced a voltage in the voltage taps, which gave a background voltage. It was easy to distinguish the voltage created by the superconductor on this background, because the voltage developed very abruptly when the current reached the value of $I_c$. It is interesting to note that the total current that the wire can carry reached 665A at 24K and 1.1T. The transport $I_c$ for the 10wt% SiC doped MgB$_2$/Fe reached 660A at 5K and 4.5T and 540A at 20K and 2T. Due to the limitation of our power source all the $I_c$ measurements were limited to the maximum 700A.

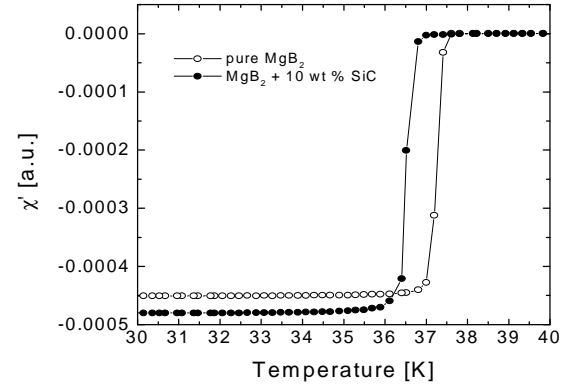

Fig. 1. Critical transition temperature ($T_c$) measured using magnetic susceptibility versus temperature for pure MgB$_2$ and 10wt% SiC doped MgB$_2$/Fe wires

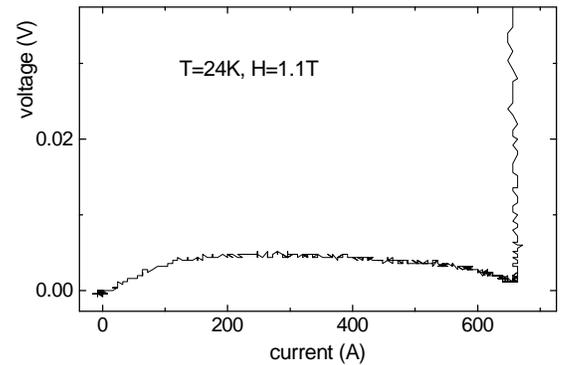

Fig. 2. I-V curve for non-doped MgB$_2$/Fe wire. $I_c$ = 665A at 24K and 1.1T.

Fig. 3 shows the $J_c(H)$ curves for the undoped and the 10wt% SiC-doped MgB$_2$ samples at 5K, 10K, and 20K. It is noted that all the $J_c(H)$ for 10wt% SiC doped MgB$_2$/Fe wire are significantly higher than the undoped sample at higher fields. The transport $J_c$ for the 10wt% doped MgB$_2$/Fe reached 133,000A/cm$^2$ at 5K and 4.5T and 108,000A/cm$^2$ at 20K and 2T. The transport $J_c$ for the 10wt% SiC doped MgB$_2$ wire increased by a factor of 6 at 5K and 9T, and 20K and 5T respectively, compared to the undoped wire. These results indicate that SiC doping strongly enhanced the flux pinning of MgB$_2$ in magnetic fields. The enhancement of pinning by SiC doping is also evident from the pinning force density versus magnetic filed as shown in Fig. 4. The volume pinning force density of 5.5GN/m$^3$ at 20K is comparable to that of NbTi at 4.2K. Although the maximum pinning force density only has a little shift to higher field, the pinning force density for the SiC doped MgB$_2$/Fe wire is clearly greater than for the undoped wire at field above 1.5T.

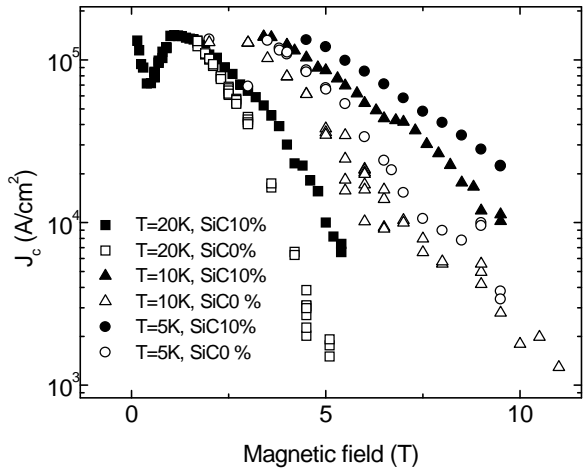

Fig. 3. The transport $J_c$ – H dependence at 5 K, 10 K and 20 K for the pure MgB$_2$/Fe and 10wt% SiC doped MgB$_2$/Fe wires

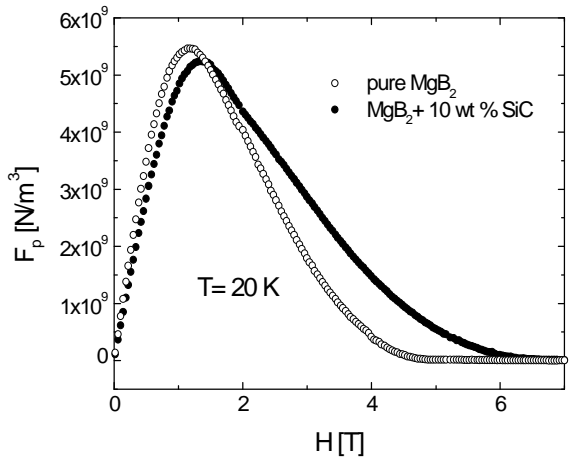

Fig. 4 Pinning force density versus magnetic filed for the undoped and 10wt% SiC doped MgB$_2$/Fe wires

the best transport $J_c$ – H performance ever reported for MgB$_2$ in any form.

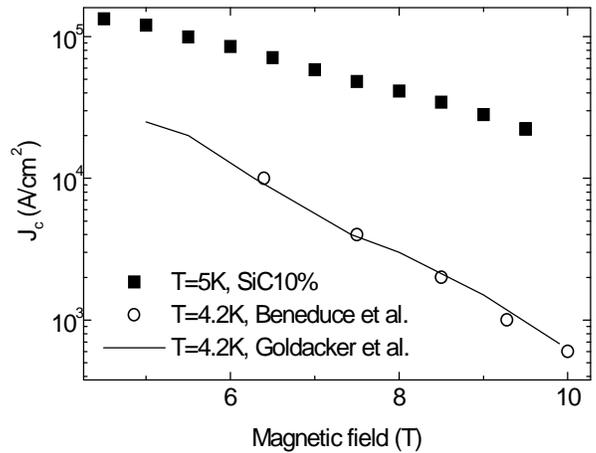

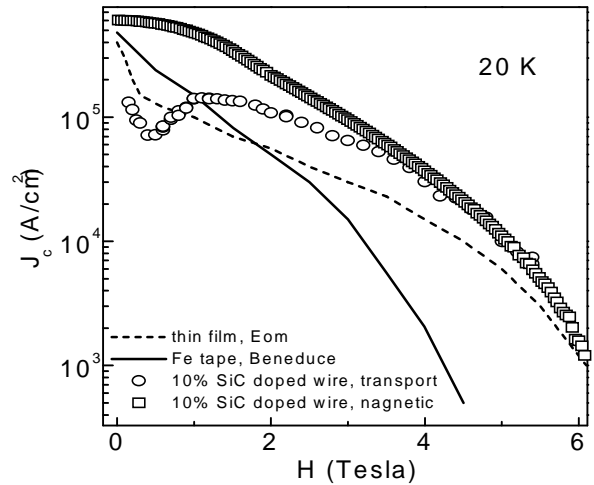

Fig. 5. A comparison of the transport $J_c$ with magnetic $J_c$ for the 10wt% SiC doped MgB$_2$/Fe wire, including the best transport $J_c$ of a strongly pinned thin film by Eom et al., [8] and Fe-sheathed MgB$_2$ tape by Goldacher et al., [3] and Beneduce et al , [14]

Fig. 5 shows the comparison of the transport $J_c$ with magnetic $J_c$. Although there is quite different voltage standard for measuring the transport $J_c$ and magnetic $J_c$, due to steep V-I characteristics, $J_c$ is expected to be similar for both methods. It is noted that the transport $J_c$ is lower than magnetic $J_c$ in low fields as the transport $J_c$ showed some type of "peak effect". The peak effect originates from the interaction between the Fe sheath and superconductor. However, the transport $J_c$ for the wires is comparable to or higher than the magnetic $J_c$ at higher fields despite the low density of the samples and percolative nature of current. Fig. 5 also shows a comparison of the transport $J_c$(H) for 10 wt% SiC doped MgB$_2$/Fe wire with the thin film [6] and the Fe-sheathed MgB$_2$ tape at 5K and 20K [3,14] reported previously. We see the $J_c$ for the 10wt% SiC doped wire is more than an order of magnitude higher than the best transport $J_c$ reported in Fe-MgB$_2$ tape at 5K and 8T and 20K and 4T respectively. It is even comparable to the strongly pinned thin film (magnetic $J_c$ for the thin film) at 20K. This is

Fig. 6 shows the $J_c$(H) versus temperature for 10wt% SiC doped wire at 1T, 2T and 4T. With SiC doping, we can achieve $J_c$ values from 50,000A/cm$^2$ to 150,000A/cm$^2$ in temperature range between 15K and 25K and field range of 2T to 5T. These results demonstrate that the nano-SiC doping into MgB$_2$/Fe wire makes a number of applications practical, including MRI, moderate magnets, magnetic windings for energy storage, magnetic separators, transformers, levitation, motors and generators. The SiC substituted MgB$_2$/Fe wire is aattractive from the economic point of view. The main cost for making MgB$_2$ conductors will be the high purity B. Furthermore, the SiC doping has already shown a significant benefit by enhancing flux pinning. It is evident that the future MgB$_2$ conductors will be made using a formula of MgB$_x$Si$_y$C$_z$ instead of pure MgB$_2$.

Fig. 7 is TEM image of Sic-doped MgB$_2$ showing a wery high density of dislocations and massive nano-meter size inclusions inside the grains 7 (a)). These inclusions are most likely the Mg$_2$Si as this is the major impurity phase picked up by the XRD analysis [15]. The EDS analyses showed that the grains consisted of Mg, B, C, Si and O. The presence of oxygen within the grain is consistent with the results obtained from a thin film with strong pinning where the ratio of Mg:B:O reached 1.0:0.9:07 [6]. All the intra-grain defects and the inclusions within the grains act as effective pinning centers, responsible for the enhanced flux pinning.

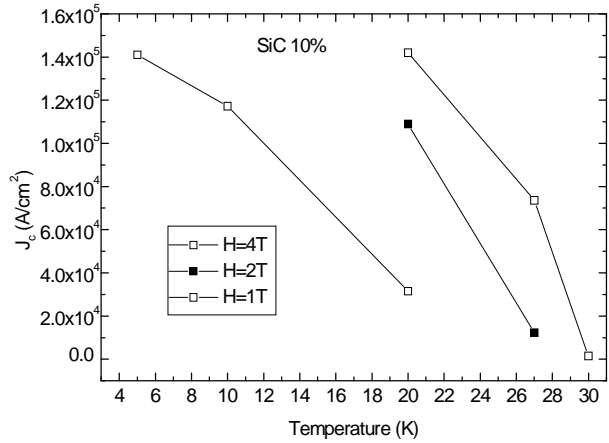

Fig. 6. J$_c$(H)versus temperature for the 10wt% SiC doped MgB$_2$ wire at 1T, 2T and 4T.

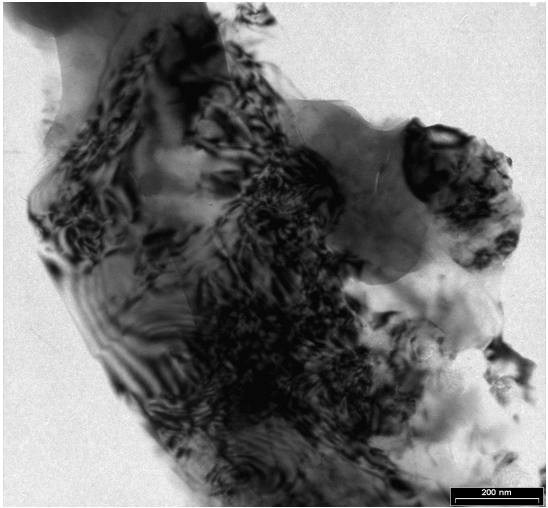

Fig. 7. TEM image for the undoped and 10wt% SiC doped MgB$_2$/Fe wires

The present study for optimization is only limited by the processing conditions, the density of the Fe-sheathed MgB$_2$ wires is still very low, only about 1.2 to 1.3 g/cm$^3$. Thus, a higher J$_c$ and better flux pinning enhancement can be achieved if the density of the samples is further improved.

## IV. CONCLUSION

In summary, we have further demonstrated that very high transport critical current and current density of Fe-sheathed MgB$_2$ wires can be achieved by a readily achievable and economically viable chemical doping with nano-SiC. J$_c$ of over 100,000A/cm$^2$ at 5K and 5T and 20K and 2T were obtained, comparable to NbTi and HTS respectively. The high performance SiC doped MgB$_2$ wires will have a great potential to replace the current market leader, Nb-Ti and HTS for many practical applications at 5K to 25K up to 5T.

ACKNOWLEDGMENT

The authors thank A. Pan, M. Ionescu, E.W. Collings, M. Sumption, M. Tomsic and R. Neale for their helpful discussion and Australian Research Council, Hyper Tech Research Inc. OH USA and Alphatech International Ltd for support.